\documentclass[twocolumn,aps,prc,showpacs]{revtex4}
\usepackage{amsmath,bm}
\usepackage{graphicx}
\usepackage{dcolumn}
\usepackage{multirow}
\begin{document}
%\begin{CJK*}{GBK}{song}
\draft
%\preprint{}
\title{Gravitation model for spatial network based on the heterogeneous node}

\author{Jiang-Hai Qian}%
\affiliation{\footnotesize College of Information, East China
Normal University, Shanghai 200062, China}
\author{ Ding-Ding Han }\thanks{Corresponding author. E-Mail: ddhan@ee.ecnu.edu.cn}
\affiliation{\footnotesize College of Information, East China
Normal University, Shanghai 200062, China}

\date{\today}
\nopagebreak

\begin{abstract}

In this paper we consider nodes in network are heterogeneous and
the link between nodes is caused by the potential dynamical demand
of the nodes. Such demand can be measured by gravitation which
increases with the heterogeneous strength of node and decreases
with the geographical distance. Based on this, we propose a new
model for spatial network from the view of gravitation. The model
is to maximize the potential dynamical demand of the whole
network, indicating the possible maximal efficiency of the network
and the highest profits that operators may gain. The model can
vary its topology by changing two parameters. A simulation for the
Chinese city airline network is completed. In the end of this
article we discuss the significance and advantage of the
heterogeneous nodes.

\end{abstract}

\maketitle

\section{Introduction}

  Since the initial studies on the small-world phenomenon were
presented by Watts and Strogatz  \cite{WS} and the scale-free
property by Barabasi and Albert  \cite{BA}, a lot of achievements
on complex network have been gotten. And our research group have
also studied some features on network \cite{hdd}. Most previous
works focus on the topological properties of the network. However
many networks are those embedded in the real space whose nodes
occupy a precise position in Euclidean space and whose links are
constrained by the geographic distance. Some typical examples are
communication networks  \cite{RA,VM},electric power grids
\cite{RIG}, transportation systems ranging from river \cite{Pitts}
to airport \cite{Amaral,Barrat,Smith}, street \cite{Crucitti},
railway and subway \cite{Latora}. In these spatial networks,
geographical factor is demonstrated to play an important role on
the network's topology.
  Very recently a few models for the spatial network have
been proposed. Some models consider both the topological and
spatially preferential attachment \cite{Waxman,Yook,xie} while
others take some optimal mechanism
\cite{Carlson,Gastner1,Gastner2}. These works provide some
guidelines in network design. On the other hand, the geographic
effects on the efficiency or traffic of the network are paid less
attention. However traffic on network is very important since it
represents the efficiency of the network. Therefore, if the
traffic between nodes can be predicted in some way, it is likely
to construct the network efficiently. Inspired by this idea and
its significance, we proposed a new spatial network model. The
model is to maximize the whole expected traffic of the network,
indicating the highest efficiency that the network may gain. The
expected traffic is measured by the gravitation. Especially, we
propose the heterogeneity of nodes and argue its significance on
the network's topology and dynamics.

\section{ Heterogeneous nodes and gravitation}

\begin{figure}
\resizebox{19.2pc}{!}{\includegraphics{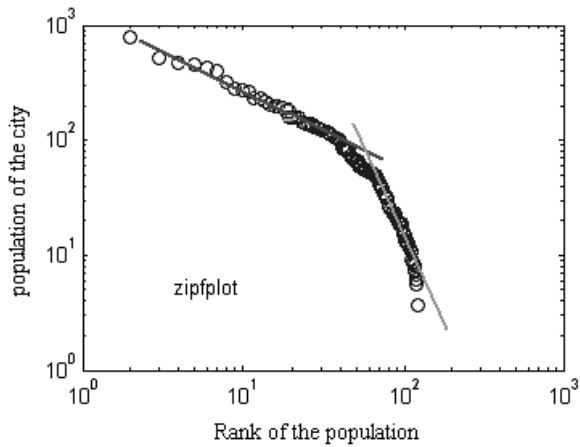}}
\caption{\footnotesize The Zipf-type plot distribution of the
population of Chinese cities. It approximately follows a
two-regime power-law distribution indicating the large variance of
the heterogeneous strength. The population information is obtained
from Ref.  \cite{China City}}
 \label{population of city}
\end{figure}

\begin{figure}
\resizebox{19.2pc}{!}{\includegraphics{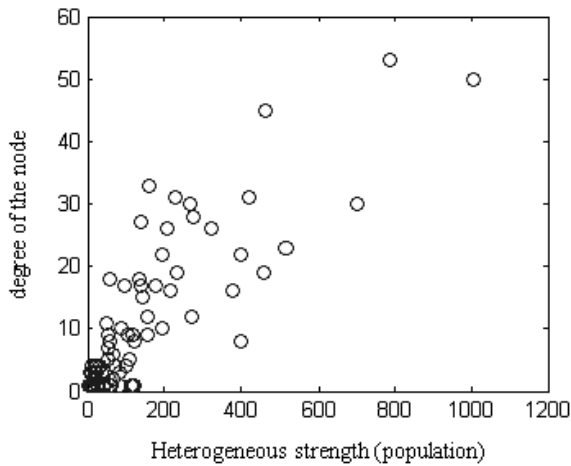}}
\caption{\footnotesize The correlation of the degree and the
population of the city in Chinese city airline network. It shows
the heterogeneous strength has a positive correlation with the
node's degree.The degree information is obtained from URL:
http://www.caac.gov.cn. }
 \label{degree-population}
\end{figure}

  In the previous studies on complex networks, nodes are usually
identical. However in the real network, situations are very
different. In the city airline network, each node (city) has
different population or economic level and such difference can be
very obvious. As is shown in Fig.~\ref{population of city}, the
population of the Chinese city follows approximately a two-regime
power-law distribution, indicating a large variance of the grade
of the city. The population can be considered as the own attribute
of node which is distinguished from others and indicates its own
grade. Such phenomenon also exists in the Internet and World Wide
Web-routers with different capacity and webs with heterogeneous
resources. Thus a node usually has its own attribute, grade or
energy indicating the heterogeneity. We define heterogeneous
strength to describe this heterogeneity, denoted by M. In
addition, we found the heterogeneous strength(i.e.the population
of city node) have positive correlation with the degree of city
node, as is shown in Fig.~\ref{degree-population}.

  Now we consider a network is composed of a group of nodes and links.
Links connect every node and realize the dynamics among them. If
two nodes with no dynamics between them, the link of the two
vertices can be thought unwanted. In a circuit network, for
example, a lead equals to disconnection if its current is 0. In
other words, it is the dynamical demand between nodes that causes
the link. Such demand is fulfilled when nodes are connected and
can be measured by the weight. However, the demand still exists
even though the vertices are not connected.

  Consider a network with n nodes and 0 edges. Every two nodes have
their demand for some information exchange. What we care is which
of these demands are the most exigent. If such information is
obtained in advance, the network will be constructed efficiently
by preferentially investing  those pairs of nodes with great
demand. To measure such demand, some of its properties should be
discussed first: (1) In a spatial network, links are constrained
by the geographic factors and the travel cost among the nodes
increases with the distance \cite{Gastner2}. So for the spatial
network, the demand of nodes is assumed to decrease with the
distance between them. (2)Since such demand can be measured by
weight after nodes are connected, we hope it has some similar
features with the real weight. In some kinds of network, such as
the airline network, the weight has the form as follows, $<w_{ij}>
\sim x_{ij}(k_ik_j)^\alpha$ \cite{AMRA,PJM}, where $k_i$ and $k_j$
are the degree of node i and j. Since degree k has a positive
correlation to heterogeneous strength M as we mentioned above, it
is reasonable to suppose such demand is proportional to
$(M_iM_j)^\alpha$.
  The above two points indicate that the potential
dynamical demand between two nodes can be described by the
following equation,
 \begin{equation}
G_{ij}=K\frac{{M_i^\alpha}{M_j^\alpha}}{D_{ij}^\gamma},
\end{equation}
Equation (1) reminds us of the Newton's gravitation equation. From
the gravitation view, the dynamical demands is considered to make
the nodes magnetize with each other and get the link when the
gravitation $G_{ij}$ is great enough. However such demand is
potential, it is a measurement of the expected traffic which is a
prediction of the real traffic.

\section{ Gravitation model for spatial network}

  We use the gravitation equation (1) to measure the potential
dynamical demand. Rewrite equation (1),
\begin{equation}
G_{ij}=K\frac{{M_i^\alpha}{M_j^\alpha}}{D_{ij}^\gamma},
\end{equation}
where K is a constant coefficient. $M_i$, $M_j$ are the
heterogeneous strength of node i and node j. Exponents $\alpha$,
$\gamma$ determine the impact of $D_{ij}$ and M on the
gravitation. If $\gamma$ takes great value, the network is
expected to be strongly constrained by the geographic distance. On
the other hand, larger value of $\alpha$ means the dominant impact
of the heterogeneous strength on the topology of the network. The
topology of the simulated network is changed by varying parameters
$\alpha$ and $\gamma$.
  The cost of constructing the network is defined as the total
number of the network's edges. It is reasonable for the airline
networks whose cost and expense are related to the total hops. But
for road network whose expense depends on the Euclidean distance,
such definition seems unconvinced. However, later discussion will
be revealed that such definition is still reasonable.
  Now suppose there are n nodes distributed on a two-dimension
plane. The heterogeneous strength and coordinates of each node are
known. By equation (2), the gravitation $G_{ij}$ of any two nodes
can be calculated. We connected preferentially those pairs of
nodes with greater gravitation $G_{ij}$ and completed such process
when the link comes to the value we preset (namely the cost). Such
process can be described as an optimal model,
\begin{eqnarray}
Max:   G = \sum_{i<j} G_{ij} k_{ij} = \sum_{i<j} K
\frac{{M_i^\alpha}{M_j^\alpha}}{D_{ij}^\gamma} k_{ij}, \nonumber \\
s.t.: \sum_{i<j} k_{ij} = \epsilon ,
\end{eqnarray}
where $k_{ij}$ is adjacency matrix element of the network. The
upper part of the equation means to take the maximum and the
bottom part means the maximum process with that constraint.
However, the above process may cause some isolated nodes. So two
more restrictions are introduced to ensure each node is connected,
\begin{eqnarray}
\sum_{i} k_{ij} \geq 1   (j=1,2,3,...,n) \nonumber \\
\sum_{j} k_{ij} \geq 1  (i=1,2,3,...,n)
\end{eqnarray}

\begin{figure} \resizebox{9.2pc}{!}{\includegraphics{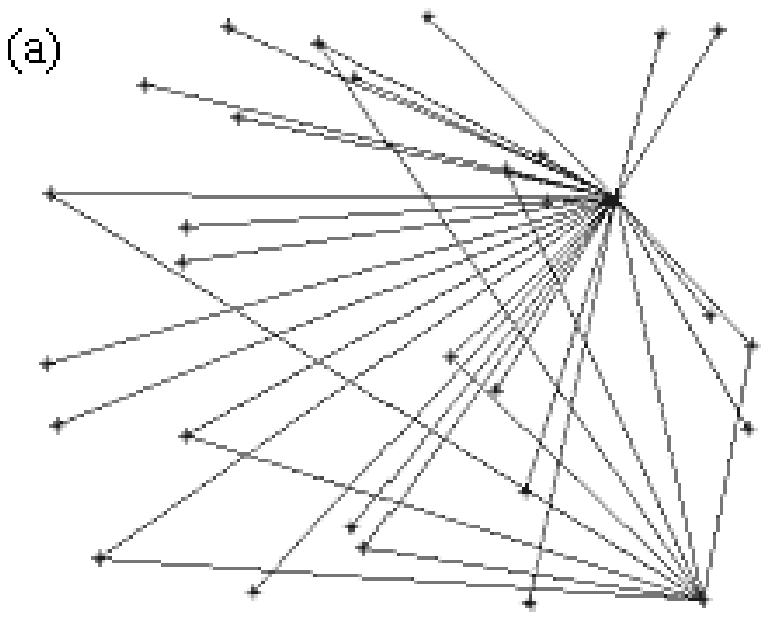}}
\resizebox{9.2pc}{!}{\includegraphics{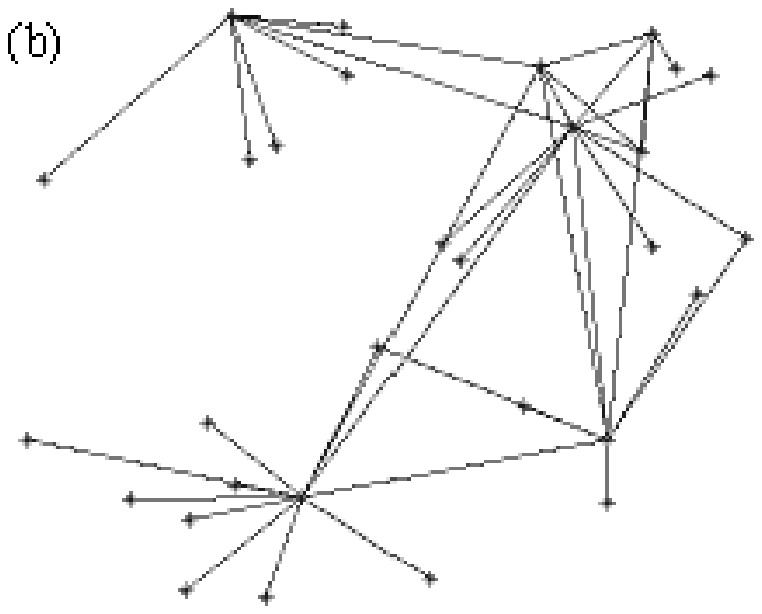}}
\resizebox{9.2pc}{!}{\includegraphics{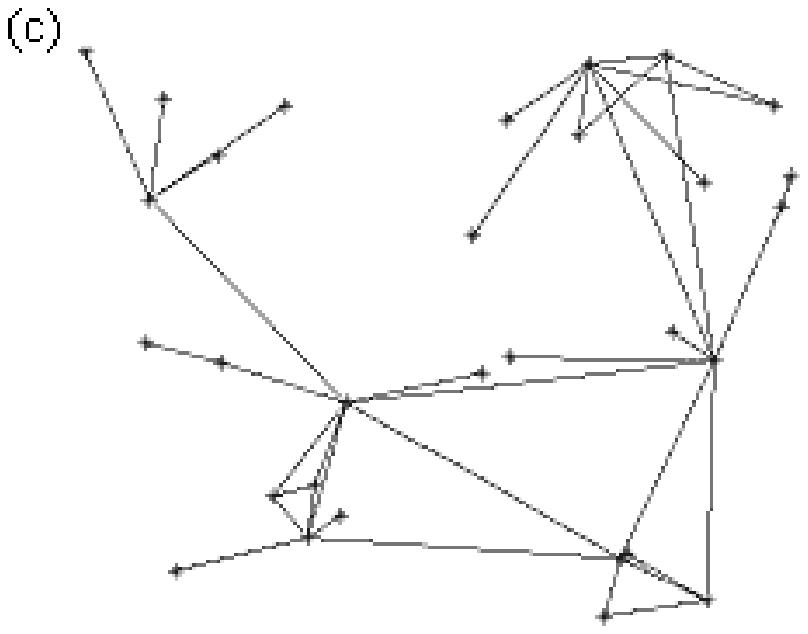}}
\resizebox{9.2pc}{!}{\includegraphics{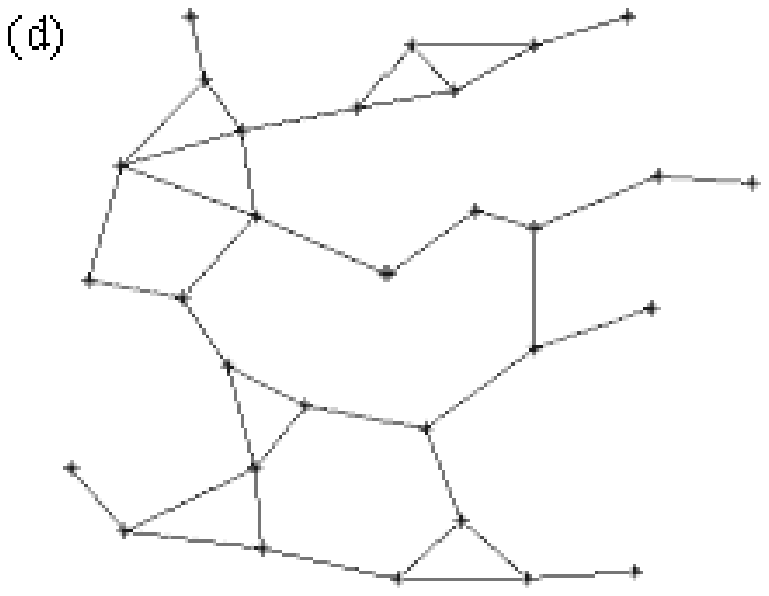}}
\caption{\footnotesize Four networks with different topology are
shown which is controlled by the parameters $\alpha$ and $\gamma$.
(a) $\alpha$ takes any value and while $\gamma$=0. The network is
dominated by two large hubs when $\gamma$=0; (b) $\alpha=1$ and
$\gamma = 1$. With the increasing of $\gamma$, the hubs become
much smaller than figure (a); (c) $\alpha$ = 1 and $\gamma$ = 2;
(d) $\alpha$ = 0, $\gamma$ takes any value. The topologies are
strongly reminiscent of airlines and roads, respectively. }
 \label{topology}
\end{figure}

  Following the method above, a network of thirty nodes is
simulated. The thirty nodes are distributed on the two-dimension
plane randomly and each node is assigned a heterogeneous strength.
Set the cost $\epsilon$ = 39 and the coefficient K = 1 (actually K
makes no difference to the model). By varying the value of
$\alpha$ and/or $\gamma$, we got four networks with different
topology as is seen in Fig.~\ref{topology}.

  For Fig.3(a), the value of $\gamma$ = 0 makes the network only rely
on the heterogeneous strength of nodes, This causes the topology
is dominated by two hubs since the nodes of great heterogeneous
strength is easy to magnetize others. Whereas with the $\gamma$
increasing, more effect of the geographic factor makes the node
tend to connect to the closer ones and weaken the hub-and -spoke
effect. When $\alpha$=1 and $\gamma$=2 (Fig.3(c)), the network
exhibits some features similar to the airline network. When
$\alpha$=0 (Fig.3(d)), the topology is entirely constrained by the
geography, which forms a two-dimensional network strongly
reminiscent of roads.
  As we have defined in the section II, the gravitation $G_{ij}$ describes
some dynamical demands or expected traffic. From the view of
operator, the significance of the equation (3) is to maximize the
whole expected traffic of the network which reflects the
efficiency of the network. In other words, our model provides a
possible way for operators to gain the highest profit. Even though
a link costs much, the operators will still earn the cost and gain
profit in short order as long as the traffic within the link is
large enough. Thus such a link will be constructed all the time.
In summary, the cost is a minor factor compared to the dynamical
demand and that is the reason we simplify the cost.

\section{Simulation for the Chinese City Airline Network}

  The gravitation model we proposed is used to simulate the Chinese airline
network  \cite{Liu}. The cities are considered as the nodes of the
network while the edges reflect the airlines. To reproduce the
real network better, we set the number of nodes n=121 and number
of links E=689. The position of each node is sited on the
two-dimensional plane, as it is shown in Fig.~\ref{geographic
distribution}. The heterogeneous strength M and the distance
$D_{ij}$ are, respectively, defined as the population of the city
and the Euclidean distance of city i and j. Selecting $\alpha$=1
and $\gamma \in$[1,2]  \cite{Chen}, they are to describe the
interactions of cities. Here we set $\gamma$ = 1.5.

\begin{figure}
\resizebox{19.2pc}{!}{\includegraphics{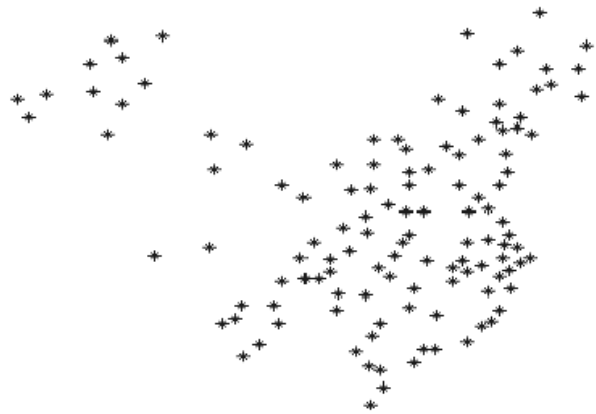}}
\caption{\footnotesize The geographic distribution of the cities
with airports. Each city denotes a node. There are totally 121
nodes in our simulation. }
 \label{geographic distribution}
\end{figure}

\begin{figure}
\resizebox{19.2pc}{!}{\includegraphics{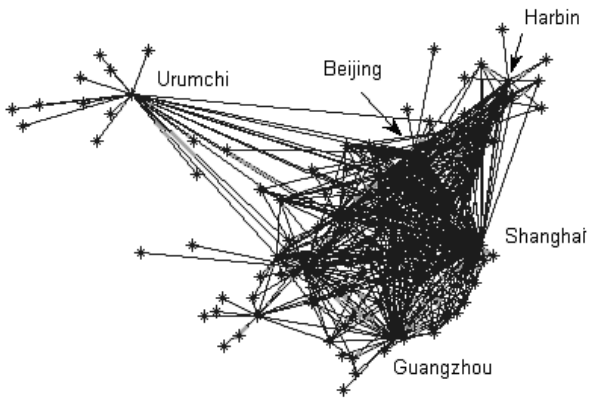}}
\caption{\footnotesize The simulation for the Chinese airline
network. Each link of two nodes reflects an airline between them.
There is totally 689 links. The cities, such as Beijing, Shanghai,
Guangzhou, Harbin and  Urumchi, exhibit hub-and-spoke phenomenon
in our simulation just as they do in the real network. }
 \label{simulation}
\end{figure}

\begin{figure}
\resizebox{19.2pc}{!}{\includegraphics{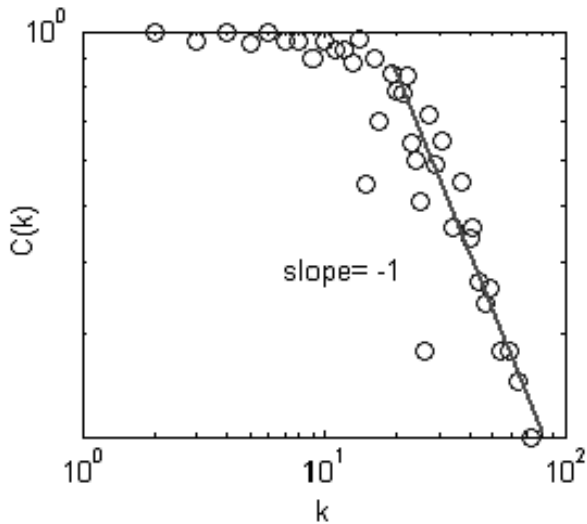}}
\caption{\footnotesize The clustering-degree distribution of the
simulation. The distribution satisfies a linear decreasing feature
with slope -1 in log-log coordinate, which indicates that the
result of simulation reproduces the hierarchy.}
 \label{clustering-degree distribution}
\end{figure}

  Fig.~\ref{simulation} shows the simulated network. Obviously, the hubs in
the real network such as Beijing, Shanghai, Guangzhou, Harbin and
Urumchi exhibit the similar hub-and-spoke phenomenon in our
simulation. In real condition, Beijing, Shanghai and Guangzhou are
the three cities with the highest degree while in our model they
are, respectively, Shanghai, Beijing and Wuhan (Guangzhou is the
fourth). The reason for such difference may be that using the
population to denote the heterogeneous strength of city is
intuitive but not be exact because the economy and the
administration factors are also important indexes for the grade of
city. In spite of this difference, we still succeed in reproducing
the every hub and their hub-and-spoken phenomenon existing in the
real network.
  The average shortest-path length L and degree-degree
correlation exponent r of the model network are calculated, where
L = 2.302, r = -0.401 while in the real network L = 2.263, r =
-0.408. Fig.~\ref{clustering-degree distribution} shows the
clustering-degree distribution. It meets $C(k) \sim k^{-1}$ which
indicates the model network exhibits the same hierarchy as the
real network does.

\begin{figure}
\resizebox{19.2pc}{!}{\includegraphics{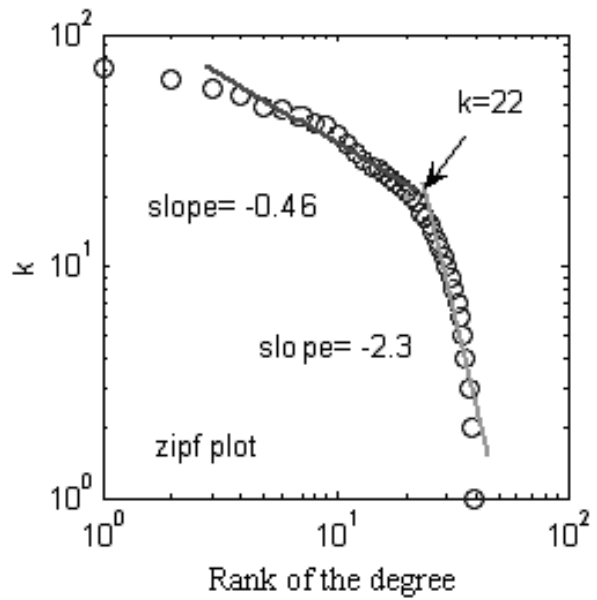}}
\caption{\footnotesize The degree distribution of the simulation.
It presents a behavior of two-regime power-law degree distribution
with the exponent$\gamma_1$ = -0.46 for the first power laws and
$\gamma_2$ = -2.3 for the second. The turning point happens at
degree k = 22. In comparison, $\gamma_1$ = -0.53, $\gamma_2$ =
-2.05 and the turning point at degree k = 20 in the Chinese City
Airline network. Both the exponents and the turning point fit the
real network well.}
 \label{degree distribution}
\end{figure}

  Fig.~\ref{degree distribution} is the degree distribution of the model network.It
satisfies the two-regime power-law distribution. This is an
interesting result not only to verify the real degree distribution
but also to produce the power-law distribution without any
preferential attachment mechanism. We are not sure if it is a
coincidence or caused by some other internal reasons, but we will
still try to offer an explanation in the following discussion.

\begin{figure}
\resizebox{19.2pc}{!}{\includegraphics{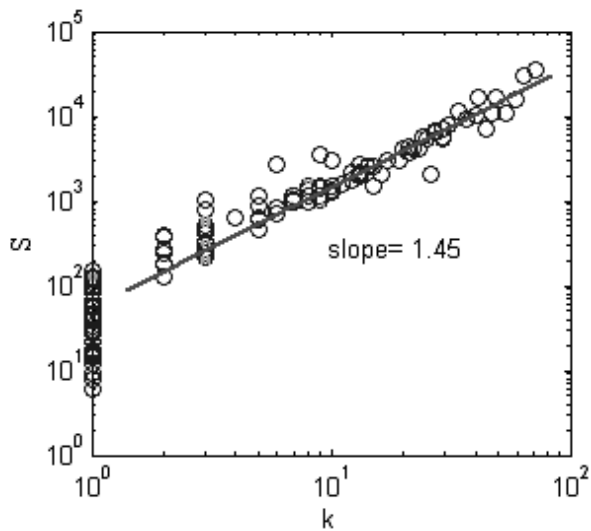}}
\caption{\footnotesize The correlation of the degree and node
strength follows $S(k)\sim k^{1.45}$, indicating the non-linearly
correlation.}
 \label{degree and node strength}
\end{figure}

  Fig.~\ref{degree and node strength} shows the correlation of the degree and node strength of
our simulation. Considering the gravitation is an effective
prediction of the real traffic, we let the gravitation of the two
nodes represent the weight of the link between them and the node
strength is the sum of such gravitation among all the links to it.
As is shown in Fig.~\ref{degree and node strength}, the node
strength increases with the degree, but does quicker than
linearly, as 1.45 power, namely satisfies $S(k)\sim k^{1.45}$.
This result reflected the real network well. In the Austrian
airline network  \cite{hdd3}and in the Chinese city airline
network it indicates that node strength increases with the degree
in a non-linearly manner.

\section{More discussions about the heterogeneous nodes}

  As we have discussed in section II, the nodes of many real
networks are usually heterogeneous. But one may ask if it is
necessary or meaningful to take this factor into consideration. We
argue the heterogeneity of nodes may be the key reason causing the
topology that we observed in many networks. Some clues are gotten
from Fig.~\ref{population of city}.The population of the city
obeys a two-regime power-law distribution which is similar to the
degree distribution of the city airline network. This reminds us
if it is the heterogeneous nodes that cause the heterogeneous
networks whose degree distribution has a large variance or in
other words, homogeneous nodes will hardly produce the hubs or
scale-free properties. An indirect evidence is that hubs in the
real network are usually those with large heterogeneous strength.
  If admitting the heterogeneity of nodes, some phenomena
such as hubs and preferential attachment can be explained from the
view of gravitation. Considering a node with large heterogeneous
strength in a spatial network,it produces a larger gravitation
field according to the gravitation view. For its extensive area of
influence, many nodes are magnetized and connected to it. So the
nodes with large heterogeneous strength usually have larger
degree, which indicates the emergence of hubs. When a new vertex
emerges, it is more possible to be magnetized by the nodes with
large heterogeneous strength in the same way, which makes it tend
to connect to these nodes (as is mentioned above, these nodes
usually have larger degree). And this just causes the phenomenon
of the preferential attachment observed in network's evolvement.
Now it may be understood why our model can produce the power-law
distribution without the preferential attachment mechanism.
Because such mechanism has been produced by gravitation although
it is not introduced designedly. Constrained by the geography,
nodes with small heterogeneous strength usually connect to the
large ones nearby while the latter are able to link with other
nodes of even larger heterogeneous strength far away and so on. It
can be imagined that the network will exhibit the hierarchy by
such organizational manner.
  Besides heterogeneity of nodes is also
related to the dynamics -- the traffic. As equation (2) shows, the
heterogeneous strength influences the gravitation $G_{ij}$ which
is considered as the expected traffic of node i and j. Although it
is only a prediction of the real traffic, Fig.~\ref{degree and
node strength} shows that such prediction verifies the real
condition. This point indicates that the heterogeneous strength of
the node is possible a bridge by which we can associate the
topology with the dynamics.

 \vspace{.5cm}

\section{Conclusion}

  In this paper, link is caused by the potential dynamical demand
among nodes which makes the nodes magnetize with each other and
its value can be estimated by gravitation. Based on this, a new
model for the spatial network is proposed. It can generate the
topology of airline and road network and can behave well in
simulating the real Chinese airline network.
  The gravitation model has its
practical meaning that follows a principle of maximizing some
expected benefit to construct and optimize the network. We argue
that cost is a minor factor despite its impact of restriction
compared with the dynamical demand of nodes.
  Heterogeneous nodes
and gravitation view are two ideas indicated by our model. We
think they are important in understanding the topology and the
dynamics of the network. However,  more demonstration studies are
essential to support this point of view.

 {}
%\end{CJK*}

\footnotesize
\begin{thebibliography}{}
\bibitem{WS} D.J. Watts and S.H. Strogatz, Nature (London) {\bf 393}, 440
(1998).
\bibitem{BA} A.-L. Barab¨¢si and R.Albert, Science {\bf286}, 509 (1999); A.-L. Barab¨¢si, R.Albert, and H.Jeong, Physica A {\bf272}, 173 (1999)

\bibitem{hdd}Ding-Ding Han, Jin-Gao Liu, Yu-Gang Ma, Xiang-Zhou Cai, Wen-Qing Shen,
 Chin. Phy. Lett. {\bf  21}, 1855 (2004); Ding-Ding Han, Jin-Gao Liu, Yu-Gang Ma, Chin. Phy. Lett., in press (2008)

\bibitem{RA}R. Pastor-Satorras, A. Vespignani,
Evolution and Structure of the Internet: A Statistical Physics Approach, Cambridge University Press, Cambridge (2004).


\bibitem{VM}V. Latora, M. Marchiori, Phys. Rev. E {\bf  71}, 015103(R) (2005).

\bibitem{RIG} R. Albert, I. Albert, G.L. Nakarado, Phys. Rev. E {\bf  69}, 025103(R) (2004).
\bibitem{Pitts} F. Pitts, The Profess. Geograph. {\bf  17}, 15 (2004).

\bibitem{Amaral}R. Guimer¨¤, S. Mossa, A. Turtschi, L.A.N. Amaral, Proc. Natl. Acad. Sci. USA {\bf  102}, 7794 (2005);
 R. Guimer¨¤, L.A.N. Amaral, Eur. Phys. J. B {\bf  38}, 381 (2004).

\bibitem{Barrat}A. Barrat, M. Barth\'elemy, R. Pastor-Satorras, A. Vespignani, Proc. Natl. Acad. Sci. USA {\bf  101}, 3747 (2004).

\bibitem{Smith}D.A. Smith, M. Timberlake, Urban Stud. {\bf  32}, 287 (1995).

\bibitem{Crucitti}  P. Crucitti, V. Latora, S. Porta, arXiv:physics/0504163


\bibitem{Latora} V. Latora, M. Marchiori, Physica A {\bf  314}, 109 (2002).
\bibitem{Waxman} B. Waxman, Routing of multipoint connections, IEEE J. Selec. Areas Commun. {\bf  6}, 1617 (1988).

\bibitem{Yook}S.-H. Yook, H. Jeong, A.-L. Barab¨¢si, Proc. Natl. Acad. Sci. USA {\bf  99}, 13382 (2002).
\bibitem{xie}Yan-Bo Xie, Tao Zhou, Wen-jie Bai, Guangrong Chen, Wei-Ke Xiao, and Bing-Hong Wang, Phys. Rev. E {\bf   75}, 036106 (2007)
\bibitem{Carlson} J.M. Carlson, J. Doyle, Phys. Rev. E {\bf  60}, 1412 (1999).
\bibitem{Gastner1}M.T.Gastner, M.E.J Newman, Eur. Phys. J. B {\bf  49}, 247-252 (2006).
\bibitem{Gastner2}M.T.Gastner, M.E.J Newman, Phys. Rev. E {\bf  74}, 016117(2006).
\bibitem{AMRA}Zhenhua Wu, Lidia A.Braunstein, Vittoria Colizza, Reuven Cohen, Shlomo Havlin, H.Eugene Stanley Phys. Rev. E {\bf  74}, 056104 (2006)
\bibitem{PJM}P.J. Macdonald, E. Almaas, A.-L. Barab¨¢si, Europhys. Lett. {\bf  72}, 308 (2005).
\bibitem{Liu}Hong-Kun Liu, Tao Zhou, Acta Phys. Sin. {\bf  56}, 0106 (2007).
\bibitem{Chen}Liu J S ,Chen Y G, Scintia Geographica Sinica {\bf  20}, 0528 (2000).
\bibitem{China City}China City Statistical Yearbook 2003 (Beijing : China Statistics Press) p16
\bibitem{hdd3}Ding-Ding Han, Jiang-hai Qian, Jin-Gao Liu, submitted to Physica A


\end{thebibliography}
\end{document}